\documentclass[12pt,oneside]{article} 
\usepackage{amsfonts,amsmath,amsxtra,amssymb,latexsym, amscd,amsthm}

\def\dis{\displaystyle}
\def\om{\omega}
\def\Om{\Omega}
\def\qq{\qquad}
\def\cen{\centerline}
\def\de{\delta}

\def\vk{\vskip0.2cm}
\def\non{\noindent}
\def\ift{\infty}
\def\ra{\rightarrow}

\def\ti{\times}
\def\f{\frac}

\def\ld{\lambda}
\def\Ld{\Lambda}
\def\pa{\partial}
\def\Ga{\Gamma}
\def\a{\alpha}

\def\ep{\varepsilon}

\begin{document}
\setlength{\baselineskip}{18pt}
\cen{\bf On the amplification of sound (acoustic phonons) by absorption of laser}
\cen{\bf radiation in cylindrical quantum wires}
\vskip0.5cm
\begin{footnotesize}
\cen{ Nguyen Quoc Hung, Nguyen Vu Nhan, Nguyen Quang Bau}
\cen{Faculty of Physics, Hanoi University of Science}
\cen{334 Nguyen Trai Str., Thanh Xuan, Ha Noi}
\end{footnotesize}
\vskip0.5cm
\non
\begin{abstract} 
Based on the quantum transport equation for the electron-phonon system, the absorption coefficient of sound (acoustic phonons) by absorption of laser radiation in cylindrical quantum wires is calculated for the case of monophoton absorption and the case of multiphoton absorption. Analytical expressions and conditions for the absorption of sound are obtained. Differences between the two cases of monophoton absorption and of multiphoton absorption are discussed; numerical computations and plots are carried out for a GaAs/GaAsAl quantum wire. The results are compared with bulk semiconductors and quantum wells.
\end{abstract}
\vskip0.5cm
\non
{\bf 1. Introduction}
\vk
The theory of amplification of sound (acoustic phonons) by absorption of laser radiation in bulk semiconductors and quantum wells has been widely studied in [1$\div$6]. In [3,4,5] the problem was restricted to bulk semiconductors with monophoton absorption and multiphoton absorption processes. In [1,2] the problem was considered for quantum wells with monophoton and multiphoton absorption processes, or in a quantizing magnetic field that does not have the restriction conditions of Ref [6]. All these authors showed that the absorption coefficient of sound can be negative in some regions of values of acoustic wave vector $q$, or it turns into the amplification coefficient of sound.
\vk
In order to continue these ideas, in this paper we study theoretically the amplification of sound by absorption of laser radiation in cylindrical quantum wires with monophoton absorption and multiphoton absorption processes. To do so, we establish quantum transport equation for phonon. Solving this equation, we obtain analytical expressions for the coefficients and the conditions for the amplification of of sound.
\vskip0.5cm
\non
{\bf 2. Quantum transport equation for acoustic phonons}
\vk
Consider a cylindrical quantum wire with radius $R$ and potential: \hfill\break 
$\left\{
\matrix
V(\vec{r})=0 && r<R\\
V(\vec{r})=\ift && r>R
\endmatrix\right.$. \\
Solving Schr\"odinger equation we obtain the wave function and the energy spectrum for electron [10]:
$$\Phi(r,\theta)=\left\{
\matrix
\dis{\f 1{\sqrt{\pi R}}\f 1{J_{n+1}(A_{n,l})}
J_n\Big(\f{A_{n,l}r}R\Big)}e^{\pm in\theta}
&& r<R\\
0 \hfill &&r>R
\endmatrix\right.\ ,\quad \ep_{n,l}(\vec{k})=\f{k^2}{2m^*}+\f{A^2_{n,l}}{2m^*R^2}.$$
$A_{n,l}$  is the $l^{th}$ root of Bessel equation of order $n:\, J_n(x);\, n,l$ are quantum numbers; $m^*$ is the effective mass of electron; $\vec{k}$ is the electron wave vector (along the wire's axis: $z$ axis).
\vk
Thus in second quantization representation, the Hamiltonian for the electron-phonon system in a quantum wire in external field can be written as:
\begin{align*}
H(t) & =\sum_{n,l,\vec{k}}\ep_{n,l}\big(\vec{k}-e\vec{A}(t)\big)a_{n,l,\vec{k}}^+a_{n,l,\vec{k}}
+\sum_{\vec{q}}\om_{\vec{q}}b_{\vec{q}}^+b_{\vec{q}}+\\
& +\sum_{n,l,n',l',\vec{k},\vec{q}}C_{n,l,n',l'}(\vec{q})a_{n,l,\vec{k}+\vec{q}}a_{n',l',\vec{k}}(b_{-\vec{q}}^++b_{\vec{q}}).\tag{1}
\end{align*}
Where $a_{n,l,\vec{k}}^+$ and $a_{n,l,\vec{k}}$ ($b_{\vec{q}}^+$ and $b_{\vec{q}}$) are the creation and annihilation operators of electron (phonon); $\vec{q}$ is the phonon wave vector; $C_{n,l,n',l'}(\vec{q})$ is the interaction coefficient of electron-acoustic phonon scattering; $\vec{A}(t)=\dis{\f 1{\Om}}\vec{E}_0\cos(\Om t)$ is the potential vector, depends on the external field. All formulae are written in units where $h=1$ and $c=1$.
\vk
Process in a similar way to Ref [1,4,5,7], from (1), we obtain the quantum transport equation for acoustic phonons in quantum wires:
\begin{align*}
& \f\pa{\pa t}\big<b_{\vec{q}}\big>_t+
i\om_{\vec{q}}\big<b_{\vec{q}}\big>_t=-\sum_{n,l,n',l'}\big|C_{n,l,n',l'}(\vec{q})\big|^2\sum_{\vec{k}}\big[n_{n,l}(\vec{k}-\vec{q})-n_{n',l'}(\vec{k})\big]\int_{-\ift}^t\big<b_{\vec{q}}\big>_{t_1}\ti \\
& \ti \sum_{\nu ,s=-\ift}J_\nu\Big(\f{\ld}{\Om}\Big)J_s\Big(\f{\ld}{\Om}\Big)\exp\left\{i\big[\ep_{n',l'}(\vec{k})-\ep_{n,l}(\vec{k}-\vec{q})\big](t_1-t)-i\nu\Om t_1+is\Om t\right\}dt_1,\tag{2}
\end{align*}
Where the symbol $\big<x\big>_t$ means the usual thermodynamics average of operator $x;\, n_{n,l}(\vec{k})$ is the distribution function of electron; $\ld =\dis{\f{e\vec{q}\vec{E}_0}{m^*\Om}}$.
\vskip0.5cm
\non
{\bf 3. The amplification of sound in quantum wires with monophoton absorption process}
\vk
Using Fourier transformation, from (2), with $\de\ra +0$, we have:
\begin{align*} 
(\om -\om_{\vec{q}})B_{\vec{q}}(\om)
& =\sum_{n,l,n',l'}\big|C_{n,l,n',l'}(\vec{q})\big|^2\sum_{\vec{k}}\big[n_{n',l'}(\vec{k})-n_{n,l}(\vec{k}-\vec{q})\big]\ti \\
& \ti\sum_{\nu ,s=-\ift}J_\nu\Big(\f{\ld}{\Om}\Big)J_s\Big(\f{\ld}{\Om}\Big)
\f{B_{\vec{q}}[\om +(s-\nu)\Om]}{[\ep_{n',l'}(\vec{k})-\ep_{n,l}(\vec{k}-\vec{q})-\om -\nu\Om -i\de]},
\end{align*}
Where $B_{\vec{q}}(\om)$ is the Fourier transformation of $\big<b_{\vec{q}}\big>_t$.
\vk
Neglecting the terms of two or higher order in the interaction coefficient $C_{n,l,n',l'}(\vec{q})$, we have the dispersion equation:
\begin{align*} 
\om -\om_{\vec{q}}
& =\sum_{n,l,n',l'}\big|C_{n,l,n',l'}(\vec{q})\big|^2\sum_{\vec{k}}\big[n_{n',l'}(\vec{k})-n_{n,l}(\vec{k}-\vec{q})\big]\ti \\
& \ti\sum_{\nu =-\ift}J_\nu^2\Big(\f{\ld}{\Om}\Big)
\f 1{[\ep_{n',l'}(\vec{k})-\ep_{n,l}(\vec{k}-\vec{q})-\om_{\vec{q}}-\nu\Om -i\de]}.
\end{align*}
From this we obtain the general formula for absorption coefficient of acoustic phonons in  quantum wires:
\begin{align*}
\a(\vec{q})
& =-\pi\sum_{n,l,n',l'}\big|C_{n,l,n',l'}(\vec{q})\big|^2\sum_{\vec{k}}\big[n_{n',l'}(\vec{k})-n_{n,l}(\vec{k}-\vec{q})\big]\ti \\
& \ti\sum_{\nu =-\ift}J_\nu^2\Big(\f{\ld}{\Om}\Big)\de\left\{\ep_{n',l'}(\vec{k})-\ep_{n,l}(\vec{k}-\vec{q})-\om_{\vec{q}}-\nu\Om\right\},\tag{3}
\end{align*}
here $\de(x)$ is the Dirac delta function.
\vk
In this section, we study the monophoton absorption process, thus the parameter in the Bessel function is small: $\ld \ll\Om$. Assume the electron is non-degenerate, we obtain the expression for the sound absorption coefficient:
\begin{align*}
a(\vec{q})
& =\f{m^*n_0\ld^2}{4q\Om^2}
\sum_{n,l,n',l'}\big|C_{n,l,n',l'}(\vec{q})\big|^2\ti \\
& \ti \exp\Big[-\f{A_{n',l'}^2}{2m^*k_BTR^2}-\f{m^*}{2k_BTq^2}(a^2+\Om^2)\Big]\Ld_{n,l,n',l'}(\vec{q},\Om),\tag{4}
\end{align*}
where $n_0$ is the electron density, $k_B$ is the Boltzmann constant, $a=\dis{\f{A_{n,l}^2-A_{n',l'}^2}{2m^*R^2}+\om_{\vec{q}}+\f{q^2}{2m^*}}$, and
\begin{align*}
A_{n,l,n',l'}(\vec{q},\Om)
& =\exp\Big(\f{\om_{\vec{q}}}{2k_BT}\Big)\biggl\{\exp\Big(-\f{m^*a\Om}{k_BTq^2}+\f{\Om}{2k_BT}\Big)sh\Big(\f{\om_{\vec{q}}+\Om}{2k_BT}\Big)+\\
& +\exp\Big(\f{m^*a\Om}{k_BTq^2}-\f{\Om}{2k_BT}\Big)sh\Big(\f{\om_{\vec{q}}-\Om}{2k_BT}\Big)\biggr\}.
\end{align*}
Due to the $\de$ function in Eq.(3), the momentum must satisfy the condition:
$$k\ge\Big|\f q2+\f{m^*}2\Big(\f{A_{n,l}^2-A_{n',l'}^2}{2m^*R^2}+\om_{\vec{q}}\Big)\pm\f{m^*}q\Om\Big|.\eqno(5)$$
From Eq.(4), we see that when $\om_{\vec{q}}\ll\Om$ the absorption coefficient is negative, or in orther word , it is the amplification coefficient of sound: $\a(\vec{q})<0$:
\begin{align*}
\a(\vec{q})
& =-\f{m^*n_0\ld^2}{2q\Om^2}
\sum_{n,l,n',l'}\big|C_{n,l,n',l'}(\vec{q})\big|^2
\exp\Big(-\f{A_{n',l'}^2}{2m^*k_BTR^2}-\f{m^*a^2}{2k_BTq^2}\Big)\ti \\
& \ti \exp\Big[\f{\om_{\vec{q}}}{2k_BT}-\f{m^*\Om^2}{2k_BTq^2}\Big]sh\Big(\f{\Om}{2k_BT}\Big)sh\Big[\f{m^*\Om}{k_BTq^2}\Big(\om_{\vec{q}}+\f{A_{n,l}^2-A_{n',l'}^2}{2m^*R^2}\Big)\Big].\tag{6}
\end{align*}
\vk
\non
{\bf 4. The amplification of sound in quantum wires with multiphoton absorption process}
\vk
Assume the parameter in Bessel function is large: $\ld\gg\Om$  and the electron is non-degenerate, use approximate formula [8,9]:
$$\sum_\nu\,J_\nu^2\Big(\f{\ld}{\Om}\Big)\de(E-\nu\Om)=\f{\theta(\ld^2-E^2)}{\pi\sqrt{\ld^2-E^2}}.\qq\theta(x)=\left\{
\matrix
1 && x>0\\
0 && x<0.
\endmatrix\right.$$
From the general formula (3), we obtain the analytical expression of the coefficient of sound:
\begin{align*}
\a(\vec{q})
& =\f{\pi^{\f 12}n_0m^*}{2q}\exp\Big(-\f{m^*\ld^2}{2k_BTq^2}\Big)
\sum_{n,l,n',l'}\big|C_{n,l,n',l'}(\vec{q})\big|^2\ti \\
& \ti\sum_{\nu =0}^\ift\f{\Ga\big(\nu +\f 12\big)}{\nu !}\Big[\chi_{n,l}\Big(\f{-q^2}{2m^*}\Big)-\chi_{n',l'}\Big(\f{q^2}{2m^*}\Big)\Big],\tag{7}
\end{align*}
where: 
\begin{align*}
\chi_{n,l}(x)
& =\exp\Big[-\f{-A_{n,l}^2}{2m^*k_BTR^2}-\f{m^*}{2q^2k_BT}\Big(\f{A_{n,l}^2-A_{n',l'}^2}{2m^*R^2}+x+\om_{\vec{q}}\Big)^2\Big]\ti \\
& \ti\Bigg(\f{\ld}{\dis{\f{A_{n,l}^2-A_{n',l'}^2}{2m^*R^2}+x+\om_{\vec{q}}}}\Bigg)^\nu I_\nu\Big[\f{m^*\ld}{q^2k_BT}\Big(\f{A_{n,l}-A_{n',l'}}{2m^*R^2}+x+\omega_{\vec q}\Big)\Big],
\end{align*}
$I_\nu(x)$ is the complex Bessel function of order $\nu$.
\vk
From the $\theta$ function above, we derive the momentum condition:
$$k<\pm\f q2+\f{A_{n,l}^2-A_{n',l'}^2}{2qR^2}+\f{m^*}q|\ld |.\eqno(8)$$
Note that if 
$$\chi_{n,l}\Big(\f{-q^2}{2m^*}\Big)<\chi_{n',l'}\Big(\f{q^2}{2m^*}\Big),\eqno(9)$$ 
then $\a(\vec{q})<0$, which means we have the amplification coefficient of sound.
\vfill\eject
\non
{\bf 5. Discussion}
\vk
From the obtained result (4), we plotted the dependence of the absorption coefficient of sound on wave vector (figure 1), on wave vector and laser frequency (figure 2), and on temperature $\beta =kT$ (figure 3) for the case of monophoton absorption in a GaAs/GaAsAl quantum wire with $R=100\overset{\circ}{A},\ m=0.067m_0,\ \hbar =1,\ c=1$. Note that the plots include both positive region and negative region (which means we have the regions of sound absorption and sound amplification). In figure 1, we can see a maximum in the amplification of sound at $q\approx 0.0004$  meV, while the maximum for the case of absorption of sound (the positive region in the graph) is obtained at $q\approx 0.0012$ meV. The graph has a similar form with the correspondent graph in quantum wells [1], except that the region of wave vector is different. Figure 2 shows the dependence of $\a$ on phonon's wave vector and laser frequency. Note that the maximum and the minimum of $\a$ do not change with laser frequency. Instead, the absorption coefficient decreases slightly with increasing laser frequency. Figure 3 is plotted for the dependence of $\a$ on temperature $\beta =kT$, with $k$ is Boltzmann constant. The laser frequency is 50 meV. Figure 3a is plotted with a value of $q\approx 0.0004$ meV (the minimum of ( in Fig.1), whereas figure 3b is plotted with a value of $q\approx 0.0012$ meV (the maximum of $\a$ in Fig.1). Both graphs show that the absorption or amplification is strongest when $\beta$ is in the interval of $0.05$ and $0.1$ meV.
\vk
The formulae (4), (7) are carried out only when the conditions (5), (8) are satisfied. That is, only phonons with proper momentum are absorbed or amplified. In the limiting case when $\om_{\vec{q}}\ll\Om$ (for monophoton absorption), $\a(\vec{q})<0$ and we have the amplification coefficient of sound, or the number of phonon is increased with time. These formulae are more complicated than the correspondent formulas in quantum wells due to the intricate dependence of the electron wave function and energy spectrum on Bessel function. With bulk semiconductors, because of the continuance of the energy spectrum, the dependence has a completely different form [3].
\vk
In the case of multiphoton absorption, the absorption coefficient and the momentum condition intricately depend on the variable (the exponential and the complex- Bessel function). The dependence on the complex - Bessel function $I_v(z), z$ varies with the intensity of the laser radiation field $\big(\ld =\dis{\f{e\vec{q}\vec{E}_0}{m^*\Om}}\big)$ showed that the amplification of sound depends on the intensity of the radiation field with the order greater than two, while the dependence in the case of monophoton absorption is of the order two. When the condition (9) is satisfied, we also have the amplification of sound. Compare these formulae with the expressions in [3] we realize the difference for both the dependence and the momentum conditions.
\vskip0.5cm
\non
{\bf 6. Conclusion}
\vk
In conclusion, we want to emphasize that:
\vk
\begin{enumerate}
\item The quantum transport equation for phonon in quantum wires was established, which has a similar form with those in quantum wells and bulk semiconductors.
\item Analytical expressions and conditions for the absorption coefficient of sound (acoustic phonons) were first obtained in the case of monophoton absorption and multiphoton absorption. In proper conditions, the absorption coefficient of sound is negative and turns out to be the amplification coefficient of sound.
\item The results are general and avoided the approximations used in [6]. The dependence of the absorption coefficients of sound is more complicated than those in the case of quantum wells and bulk semiconductors.
\end{enumerate}
\vskip0.5cm
\non
{\bf Reference}
\vk
\begin{enumerate}
\item[{[1]}] Nguyen Quang Bau, Vu Thanh Tam, Nguyen Vu Nhan, Military science and technology magazine, No 24, 3 (1998) 38.
\item[{[2]}]  Nguyen Quang Bau, Nguyen Vu Nhan, Nguyen Manh Trinh. Proceedings of IWOMS '99, Hanoi 1999, 869.
\item[{[3]}]  Nguyen Quang Bau, Nguyen Vu Nhan, Chhoumm Navy, VNU. Journal of science, Nat.sci., 15 (1999) 1.
\item[{[4]}]  E.M. Epstein, Radio Physics, 18 (1975) 785.
\item[{[5]}]  E.M. Epstein, Lett. JETP, 13 (1971) 511.
\item[{[6]}] Peiji Zhao, Phys. Rev., B49 (1994) 13589.
\item[{[7]}] Nguyen Quang Bau, Nguyen Van Huong, J. Science of HSU, Physics, 3 (1990) 8.
\item[{[8]}] Nguyen Hong Son, Shmelev G. M, Epstein E. M, Izv. VUZov USSR, Physics, 5 (1984) 19.
\item[{[9]}]  L. Sholimal, Tunnel effects in semiconductors and applications, Moscow, (1974).
\item[{[10]}]  Gold and Ghazali, Phys. Rev. B41 (1990) 8318.
\end{enumerate}

\vskip0.5cm
\non
{\bf Figure}
\vk
\begin{enumerate}
\item[{Fig. 1}] Dependence of absorption coefficient of sound on wavevector with $\beta =0.08$ meV, $\Om =50$ meV, $\hbar =1, c=1$. 
\item[{Fig. 2}] Dependence of absorption coefficient on wave vector and laser frequency, with $\beta =0.08$ meV, $\hbar =1, c=1$.
\item[{Fig. 3}] Dependence of absorption coefficient on temperature. Figure 3a is plotted at $q=0.0004$ meV and figure 3b at $q=0.0012$ meV, $\Om =50$ meV, $\hbar =1,\  c=1$.
\end{enumerate}

\end{document}
\vskip0.5cm
Figure 1: Dependence of absorption coefficient of sound on wavevector with $\beta =0.08$ meV, $\Om =50$ meV, $\hbar =1, c=1$. 
\vk
Figure 2: Dependence of absorption coefficient on wave vector and laser frequency, with $\beta =0.08$ meV, $\hbar =1, c=1$.
\vk
\cen{Figure 3a\hskip6cm Figure 3b}
\vk
Figure 3: Dependence of absorption coefficient on temperature. Figure 3a is plotted at $q=0.0004$ meV and figure 3b at $q=0.0012$ meV, $\Om =50$ meV, $\hbar =1,\  c=1$.
\vskip0.5cm